\documentclass[aps,prl,showpacs,twocolumn,superscriptaddress,letterpaper]{revtex4}
\usepackage[pdftex]{graphicx}
\usepackage{amsmath}

\def\bea{\begin{eqnarray}}
\def\eea{\end{eqnarray}}
\def\ben{\begin{equation}}
\def\een{\end{equation}}
\def\benu{\begin{enumerate}}
\def\enu{\end{enumerate}}

\def\la{{\langle\, }}
\def\ra{{\,\rangle }}
\def\p{\,|\,}

\def\n{n}

\def\sss{\scriptscriptstyle\rm}

\def\1var{(\bx_1...\bx\N)}

\def\br{{\bf r}}

\def\bx{{\br t}}

\def\N{_{\sss N}}

\def\ext{_{\rm ext}}

\newcommand{\ee}{{\rm ee}}

\def\sph_int{ {\int d^3 r}}

\begin{document}

\title{Time-dependent natural orbitals and occupation numbers}
\author{H. Appel}
\email[Electronic address:\;]{appel@physik.fu-berlin.de}
\affiliation{Fachbereich Physik, Freie Universit\"{a}t Berlin, Arnimallee 14, D-14195 Berlin, Germany}
\affiliation{Fritz-Haber-Institut der Max-Planck-Gesellschaft, Faradayweg 4-6, D-14195 Berlin, Germany}
\affiliation{European Theoretical Spectroscopy Facility (ETSF)}

\author{E.K.U. Gross}
\affiliation{Fachbereich Physik, Freie Universit\"{a}t Berlin, Arnimallee 14, D-14195 Berlin, Germany}
\affiliation{European Theoretical Spectroscopy Facility (ETSF)}

\date{\today}
\begin{abstract} 
We report equations of motion for the occupation 
numbers of natural spin orbitals and show that adiabatic extensions of
common functionals 
employed in ground-state reduced-density-matrix-functional theory
have the shortcoming of leading always to occupation numbers which 
are {\em independent of time}. 
We illustrate the exact time-dependence of the natural spin orbitals 
and occupation numbers for the case of electron-ion scattering and 
for atoms in strong laser fields. In the latter case, we observe
strong variations of the occupation numbers in time.
\end{abstract}

\pacs{31.15.Ew, 31.70.Hq, 34.80.-i, 32.80.Qk}

\maketitle


The description of quantum many-body systems out of equilibrium has become 
an important research topic in nuclear, plasma and 
condensed-matter physics. The common interest of different fields in non-equilibrium 
quantum evolution is mainly driven by experimental and technological
progress which raises questions such as how many-body systems evolve on transient or
non-adiabatic time-scales, how they thermalize or which kind of transport phenomena 
are to be expected in extreme environments.

The Kadanoff-Baym equations \cite{KB62,D84,B98} provide a rigorous basis to 
investigate the dynamics of non-equilibrium many-body systems. 
Although the equations are known for some decades, so 
far only ab-initio solutions for very small atomic or molecular systems under special 
assumptions have been achieved \cite{DL07,TR08}.
For a more comprehensive ab-initio treatment of non-equilibrium systems with many degrees
of freedom they are still out of scope of present day computing facilities.

An alternative for the study of non-equilibrium processes in many-body systems is 
provided by time-dependent density-functional theory (TDDFT)~\cite{MUN06}. 
TDDFT is currently the method of choice for electronic systems out of equilibrium
because it generally achieves decent accuracy at affordable computational cost.
All physical observables are in principle functionals of the time-dependent density \cite{RG84, L99}.
However, in practice it is sometimes rather cumbersome to find approximate expressions
for observables of interest. Prominent examples are double and above-threshold ionization of
atoms and molecules in strong laser fields, where no accurate functionals for the 
observables are known \cite{LL98}. 

Recently, there has been renewed interest in reduced-density-matrix-functional 
theory (RDMFT)~\cite{GU98, BB02}. RDMFT is a promising candidate to treat long standing
problems present in traditional density-functional theory (DFT). 
RDMFT describes very accurately the cohesion and dissociation of diatomic molecules
which presents a difficult problem for DFT methods due to the importance of 
static correlation \cite{GPB05}. The theory has also been successfully applied to the calculation 
of the fundamental gap \cite{HLA05}. 

In this work we consider the description of non-equilibrium many-particle systems
in terms of time-dependent reduced-density matrices. 
These density matrices
have the potential to capture the physics of strong correlations \cite{SDLG08}
in the time-dependent case and, furthermore, observables can be more easily 
extracted from density matrices than from the density alone.


The time-evolution of the reduced density matrices is given by
the Bogoliubov-Born-Green-Kirkwood-Yvon
(BBGKY) hierarchy of equations of motion \cite{B61}.
The first equation of the hierarchy has the form
\begin{equation}
\begin{split}
\label{eq:bbgky1}
i\partial_t \gamma_1(&11';t)= \left( \hat{h}_0(1;t) - \hat{h}_0(1';t) \right) \gamma_1(11';t) \\
            & + \int \left( v_\ee(12) - v_\ee(1'2) \right) \gamma_2(121'2';t)|_{2'=2} \,d2,
\end{split}
\end{equation}
where the bare single particle Hamiltonian is given by
$\hat{h}_0(1;t) = -\nabla^2_1/2 + v\ext(1)$ and $v_\ee(12)$ 
denotes the particle-particle interaction. The coordinates are written as
combined space-spin variables $1 = (\br_1,\sigma_1)$ and we use 
$\int d1 = \sum_{\sigma_1} \int d\br_1$. In general, similar equations in the BBGKY
hierarchy connect the time-evolution of the N-th order 
reduced density matrix $\gamma_N$ to the density matrix of order N+1. 
Since $2N$ cordinates appear in the equation of order N, the propagation
of the complete hierarchy is even more involved than solving the 
underlying time-dependent Schr\"odinger equation (TDSE). Hence, truncations of 
the hierarchy are performed in practice. To close the
system of evolution equations at a given order N it is therefore necessary 
to express the matrix of order N+1 in terms of matrices with order less or
equal to N. This is the central point where approximations enter in a time-dependent
reduced density matrix theory.
Note, that in the case of two-body
interactions $v_\ee(12)$ it is sufficient for the calculation of the ground-state energy 
to know the diagonal of the two-body reduced density matrix $\gamma_2(1212;t=0)$. 
As can be seen directly from Eq.~(\ref{eq:bbgky1}), this is in contrast to the time-dependent 
case, where also off-diagonal matrix elements $\gamma_2(121'2;t\ge 0)$ enter the equation
of motion.\\

By diagonalizing the one-body reduced-density matrix $\gamma_1(11';t)$ at each 
fixed point in time one obtains eigenvectors $\varphi_j(1;t)$ and eigenvalues 
$n_j(t)$ which we term, in analogy to static RDMFT, time-dependent natural 
orbitals and time-dependent occupation numbers, respectively \cite{PGB07}. The spectral 
representation of the one-body matrix then takes the form
\begin{equation}
\label{eq:rdm1}
\gamma_1(11';t) = \sum_{k} n_k(t) \varphi_k(1;t)\varphi^*_k(1';t).
\end{equation}
Since the eigenvectors $\varphi_j(1;t)$ form a complete set at each
point in time, we can also express the two-body reduced matrix in the basis
of natural orbitals of the one-body matrix
\begin{eqnarray}
\label{eq:rdm12}
\gamma_2(12,1'2';t) &=& \sum_{ijkl} \gamma_{2,ijkl}(t)\times \\
&& \varphi_i(1;t)\varphi_j(2;t)\varphi^*_k(1';t)\varphi^*_l(2';t). \nonumber
\end{eqnarray}
With Eq.~(\ref{eq:rdm1}) the time-dependent natural orbitals and occupation numbers have been
introduced by a diagonalization procedure.
It is now interesting to observe that the occupation numbers obey evolution
equations which have the following form
\begin{equation}
\label{eq:eomoccgamma}
i\, \dot n_k(t) = \sum_{ijl} \gamma_{2,ijkl}(t) \la ij\p v_\ee\p kl \ra(t) - c.c.
\end{equation}
where we have used $\la ij\p v_\ee\p kl \ra(t)$ as shorthand for the coulomb integrals
\begin{equation}
\label{eq:coulomb}
\int  \varphi_i(1;t)  \varphi_j(2;t) v_\ee(12)  \varphi^*_k(1;t)\varphi^*_l(2;t)\,d1d2.
\end{equation}
Next, we separate the coefficients $\gamma_{2,ijkl}(t)$ into a mean-field and 
a cummulant part $\lambda_{2,ijkl}(t)$ \cite{KM99}
\begin{equation}
\label{eq:cumulant}
\gamma_{2,ijkl}(t) = n_i(t)n_j(t)(\delta_{ik}\delta_{jl} - \delta_{il}\delta_{jk}) + \lambda_{2,ijkl}(t).
\end{equation}
Inserting (\ref{eq:cumulant}) into the equation of motion for the occupation numbers (\ref{eq:eomoccgamma})
one finds, that Hartree and exchange-like contributions
cancel. 
Consequently, we obtain that only the remaining {\em cumulant part} of the 
two-body reduced density matrix determines the time-evolution of the occupation numbers
\begin{eqnarray}
\label{eq:eomocc}
i\, \dot n_k(t) &=& \sum_{ijl} \lambda_{2,ijkl}(t) \la ij\p v_\ee\p kl \ra(t) - c.c.
\end{eqnarray}

For the closure of the BBGKY hierarchy up to level N=1 an approximation of the
two-body reduced density matrix in terms of the one-body matrix is required.
One might be tempted to extend common functionals used in static RDMFT to the
time-dependent case in a spirit similar to the adiabatic local
density approximation of TDDFT \cite{TGK08}, i.e. for a slowly varying time-dependence in 
the Hamiltonian the ground-state functional is evaluated at the time-dependent 
density. In the case of time-dependent reduced density matrices 
this corresponds to replacing the ground-state 
occupation numbers by their time-dependent counterparts. Although this replacement
is rather ad hoc, the hope would be that such functionals perform also reasonably well 
in non-adiabatic situations.


The functional form of most commonly used ground-state functionals in RDMFT,
written in the basis of the natural orbitals, can be summarized with the following 
expression
\begin{equation}
\label{eq:rdmftfunc}
\gamma_{2,ijkl} = f_{ijkl}\delta_{ik}\delta_{jl}-g_{ijkl}\delta_{il}\delta_{jk},
\end{equation} 
which contains Hartree ($\delta_{ik}\delta_{jl}$) and exchange-like ($\delta_{il}\delta_{jk}$)
terms.
As example, for the M\"uller functional \cite{M84} we have $f_{ijkl} = 1/2\, n_i\,n_j$,
$g_{ijkl} = 1/2\, \sqrt{n_i\,\n_j}$, the self-interaction corrected functional of 
Goedecker and Umrigar \cite{GU98} reads $f_{ijkl} = 1/2\, (n_i\,n_j -n^2_i\delta_{ij}\delta_{ik}\delta_{il})$,
$g_{ijkl} = 1/2\, (\sqrt{n_i\,\n_j} -n_i\delta_{ij}\delta_{ik}\delta_{il})$ and the BBC1 functional of 
Baerends et al. \cite{GPB05} has the form $f_{ijkl} = 1/2\, n_i\,n_j$, 
$g_{ijkl} = \sqrt{n_i\,\n_j}(1/2\, - \delta_{il}\delta_{jk}(1-\delta_{ij})\Theta(i-N-\epsilon)\Theta(j-N-\epsilon))$,
where $\Theta$ denotes the usual Heaviside step function and $0 < \epsilon < 1$.
In a similar fashion the BBC2/BBC3 functionals 
can be written in the form of Eq. (\ref{eq:rdmftfunc}).
Note, that all functionals have the symmetry $g_{ijkl}=g_{jilk}$ and all matrices
$f_{ijkl}$, $g_{ijkl}$ are real valued.

Replacing the static occupation numbers which appear in Eq. (\ref{eq:rdmftfunc}) 
by their time-dependent counterparts and inserting 
this approximation for the two-body matrix into the equation of motion for the 
occupation numbers (\ref{eq:eomoccgamma}) we obtain
\begin{equation}
\begin{split}
\label{eq:zerorhs}
i\, \dot n_k(t) &= \sum_{j} (f_{kjjk}(t) - f^*_{kjjk}(t)) \la kj\p v_\ee\p kj \ra(t) \\
	        &\:+ \sum_{j} (g^*_{jkkj}(t) - g_{jkkj}(t)) \la jk\p v_\ee\p kj \ra(t),
\end{split}
\end{equation}
which shows that all functionals of the form (\ref{eq:rdmftfunc}) with real valued 
matrices $f_{ijkl}$, $g_{ijkl}$ cause a zero right hand side in (\ref{eq:zerorhs}). 
Hence, if this class of approximations
is used for the time-evolution of the one-body matrix $\gamma_1$, the occupation 
numbers stay {\em constant in time}. This is a severe shortcoming of an adiabatic
extension of present functionals of static RDMFT which needs to be addressed in the development 
of future functionals.
Possible functional forms that lead to a non-vanishing right hand side in (\ref{eq:zerorhs})
would be $\gamma_{2, ijkl}(t) = g( f_{ik}(t)\,f_{jl}(t) - f_{il}(t)\,f_{jk}(t) )$
or $\gamma_{2, ijkl}(t) = g( (f_{ij}(t) - f_{ji}(t))(f_{kl}(t) - f_{lk}(t)) )$,
where $f_{ij}$ is a non-diagonal real-valued matrix and $g$ some Taylor-expandable 
function. Alternatively, functionals with complex valued matrices could be employed.


\begin{figure}[t]
  {\par\centering
  \resizebox*{0.75\columnwidth}{!}{\includegraphics{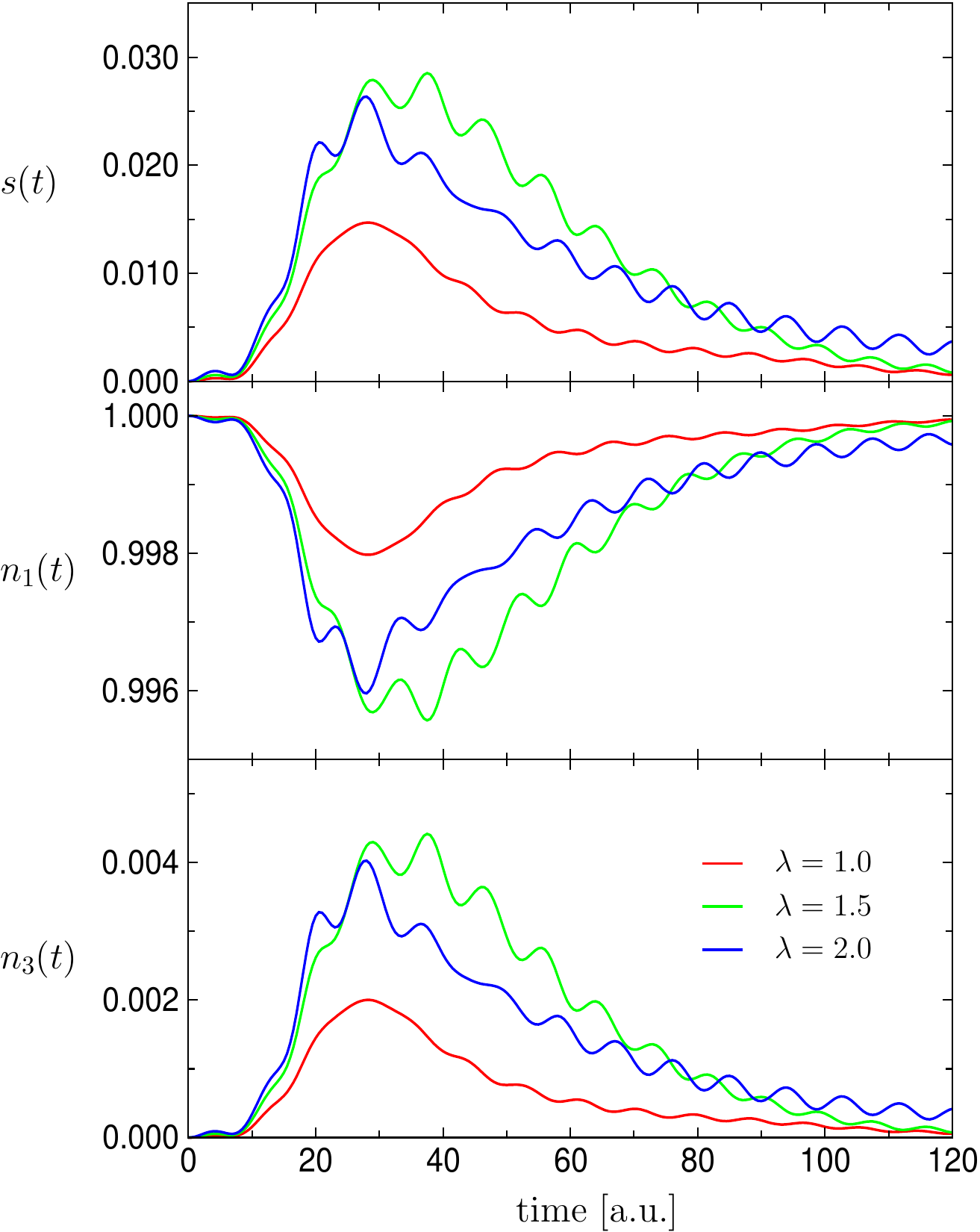}} 
  \put(-142,216){(a)}
  \put(-142,97){(b)}
  \put(-142,74){(c)}
  \par}
  \caption{(a) Correlation entropy of e-He$^+$ scattering for different
  interaction strengths $\lambda$. 
  (b), (c) Occupation numbers for the largest natural orbitals, respectively.
  \label{fig:scattocc}}
\end{figure}

How do occupation numbers actually change in time?
In the discussion above we have seen that present functionals of static RDMFT would lead 
to time-independent occupation numbers. Is this a sensible approximation?
To date the time-dependence of occupation numbers and natural orbitals is totally unknown.
Take as an example the He atom and consider
a transition from the ground state (where the two 1s natural spin orbitals have occupation 
numbers close to $1$) to the first excited state 
(where the 1s and the 2p spin orbitals have spin-degenerate occupation numbers close to 1/2) 
by virtue of an optimized laser pulse. How do occupation numbers 
and natural orbitals evolve during the transition? At first sight two extreme 
scenarios are conceivable for this case: 
(i) the occupation numbers stay nearly constant in time and only the 
orbitals change as follows: 1s $\to$ 1s, 1s $\to$ 2p;  or the opposite scenario: 
(ii) the orbitals stay close to the initial orbitals and only the occupation 
numbers change to achieve the (1s,2p) final state. Particularly interesting is the
fact that this is a transition from a weakly correlated to a strongly
correlated state (the latter being a superposition of two determinants).

To investigate the exact time-dependence of natural orbitals and occupation
numbers we consider two prototypical cases: (i) e-He$^+$ scattering 
and (ii) the above mentioned transition of the Helium atom from the ground state to the 
first excited state induced by an optimized short laser pulse. In both cases we limit 
ourselves to reduced dimensionality so that
the full time-dependent many-body Schr\"odinger equation can be solved 
numerically. From the exact time-dependent many-body wave function we compute the 
one-body reduced density matrix and by diagonalizing the matrix at each point
in time we obtain the natural orbitals and occupation numbers.
We evolve both systems (i) and (ii) under a Soft-Coulomb Hamiltonian \cite{SE91}
of the form
\begin{eqnarray}
\hat H_\lambda &=& \frac{\hat{p}_1^2}{2} + \frac{\hat{p}_2^2}{2}
    \,-\,\frac{2}{\sqrt{\hat{x}_1^2+1}}
    \,-\,\frac{2}{\sqrt{\hat{x}_2^2+1}}  \\ \nonumber
& & \,+\,\frac{\lambda}{\sqrt{(\hat{x}_1-\hat{x}_2)^2+1}}. \\ \nonumber
\end{eqnarray}
To vary the degree of correlation in the many-body wave function we introduce a coupling
constant $\lambda$ in the Hamiltonian which controls the strength of the electron-electron
interaction.

\begin{figure}
  \begin{minipage}[c]{0.85\columnwidth}
  {\par\centering
  \resizebox*{1.1\textwidth}{!}{
  \includegraphics{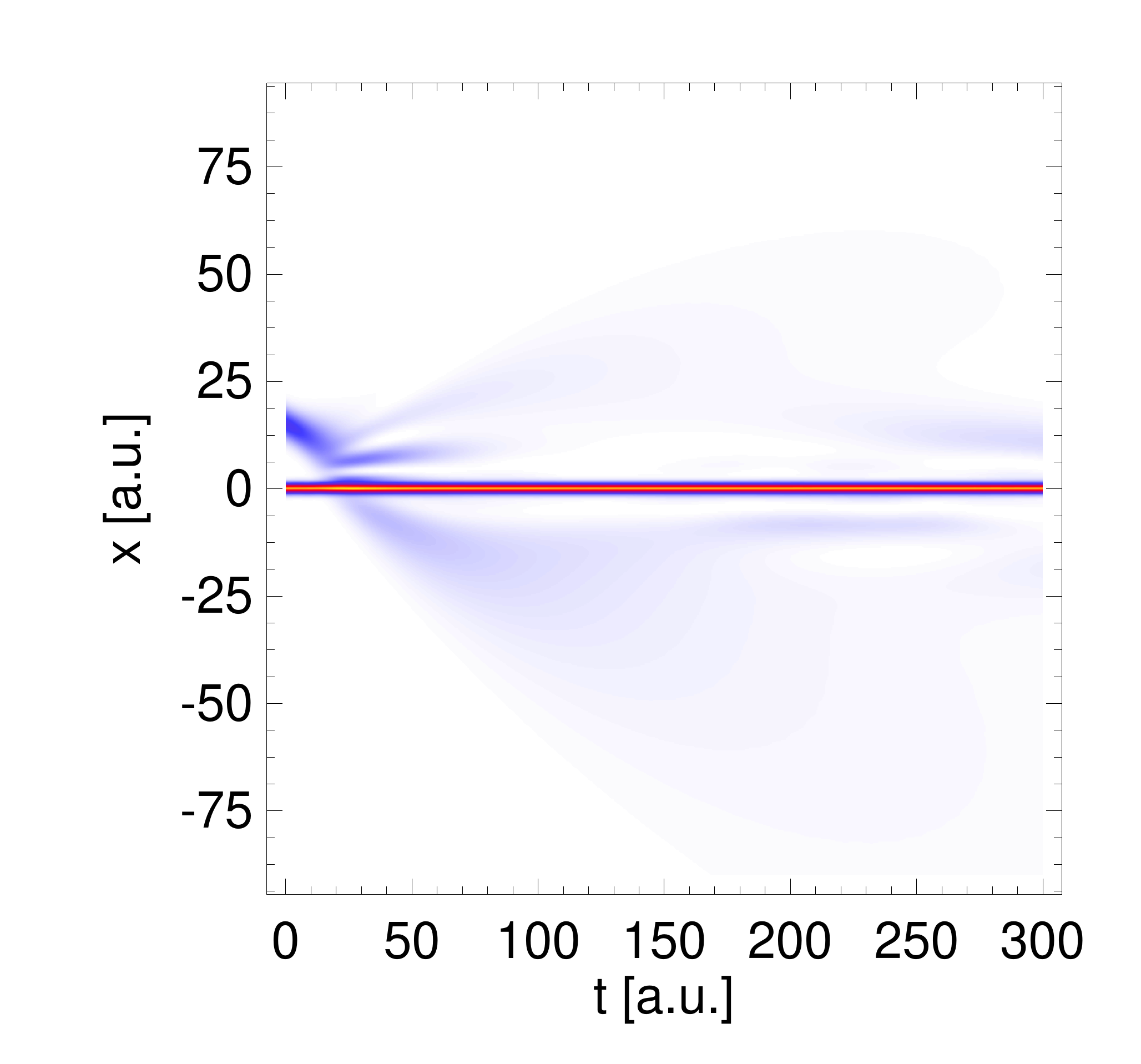}
  \includegraphics{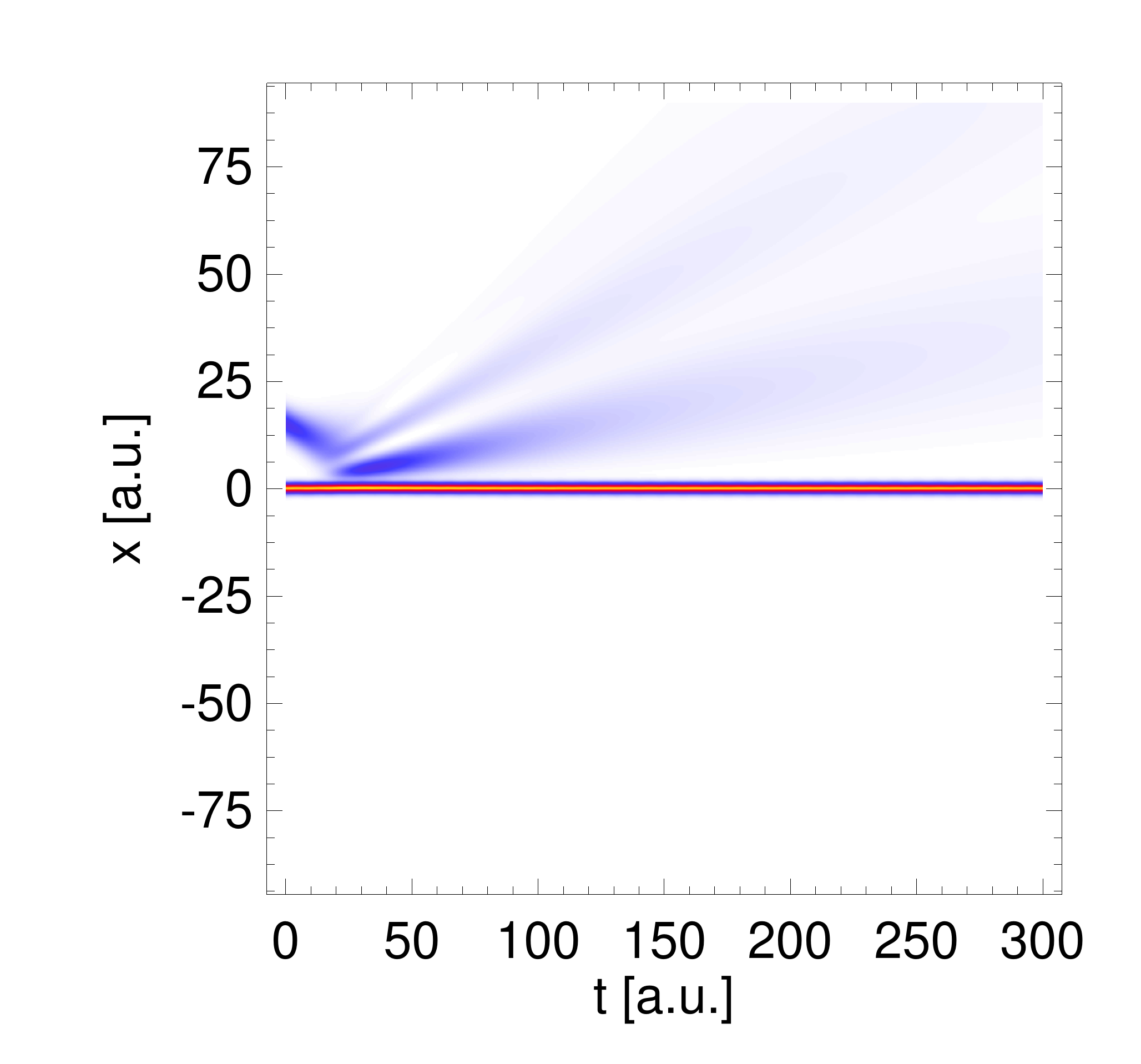}}
  \put(-230,90){(a)}
  \put(-115,90){(b)}
  \par}
  \end{minipage}
  \caption[Space-time plot of the electron density for e$^-$~--~He$^+$ scattering.]
           {Space-time plot of the electron density for e$^-$~--~He$^+$ scattering. 
           (a) $\lambda=1.0$, (b) $\lambda=2.0$.
  \label{fig:nso}}
\end{figure}

Ref. \cite{Z00} introduced the correlation entropy
\begin{equation}
s(\{n_k\}) := - \frac{1}{N}\,Tr \gamma_1 \ln \gamma_1 = - \frac{1}{N}\, \sum_k n_k \ln n_k
\end{equation}
as a measure for the correlation which is contained in the one-body reduced
density matrix. The correlation entropy $s$ defined in this way is identical to zero for 
noninteracting particles and grows with increasing correlation in the system. 
In our time-propagations we calculate $s(\{n_k(t)\})$ as function of time to monitor the
degree of correlation which is present in the many-body wave function at a 
given instant in time.

\begin{figure}[t]
  {\par\centering
  \resizebox*{0.75\columnwidth}{!}{\includegraphics{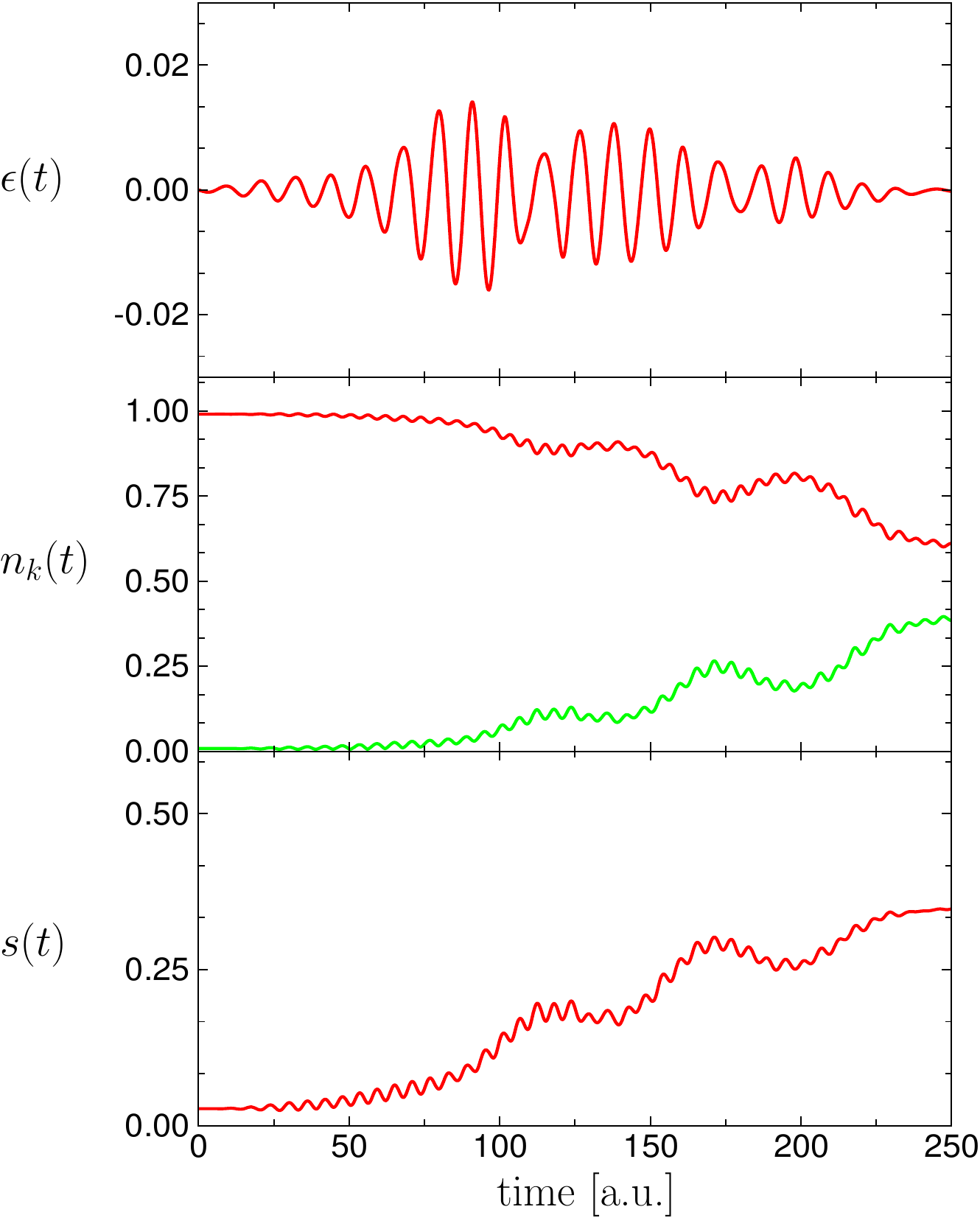}} 
  \put(-142,216){(a)}
  \put(-142,97){(b)}
  \put(-142,74){(c)}
  \par}
  \caption{(a) Optimal laser pulse for the transition of the Helium atom
  from the ground state to the first excited singlet state. The overlap of the propagated
  wave-function and the target state was 98.59\% after 12 OCT iterations.
  (b) The two largest occupation numbers and (c) the correlation entropy as function
  of time.
  \label{fig:octocc}}
\end{figure}

Case (i) - {\em electron-ion scattering}. For the initial state of the propagation we 
consider an antisymmetric spin singlet product wave function formed from a Gaussian wave  
packet $\psi(x) = \exp(-(x-x_0)^2/\sigma^2)\exp(-k_0 x)$ and the ground state $\phi_0(x)$ of the
He$^+$-ion. For all calculations we place the
packet initially at a distance of $x_0=15$ a.u. away from the ionic core and give
the scattering electron a momentum of $k_0 = 0.3$ a.u. which is pointing towards the ion.
In Fig.~\ref{fig:scattocc} we plot the occupation numbers and the 
correlation entropy as function of time for different values of the electron-electron
interaction strength $\lambda$. After about $t=10$ a.u. the wave packet has approached the ionic core and 
is passing the atomic nucleus. During this time the degree of correlation
in the wave function is enhanced. After the collision the transmitted and reflected 
waves are leaving the ionic center (cf. Fig.~\ref{fig:nso}) and the correlation entropy 
starts to decrease. As indicated by the correlation entropy for long times after the 
scattering event, the many-body wave-function is again
well represented by a single Slater determinant. The occupation numbers deviate most strongly from their determinantal
values (0 or 1) for $\lambda=1.5$. For $\lambda=2.0$ the electron-electron repulsion is
already so strong that the incident wave packet is mainly reflected back. 

\begin{figure}
  {\par\centering
  \resizebox*{0.60\columnwidth}{!}{\includegraphics{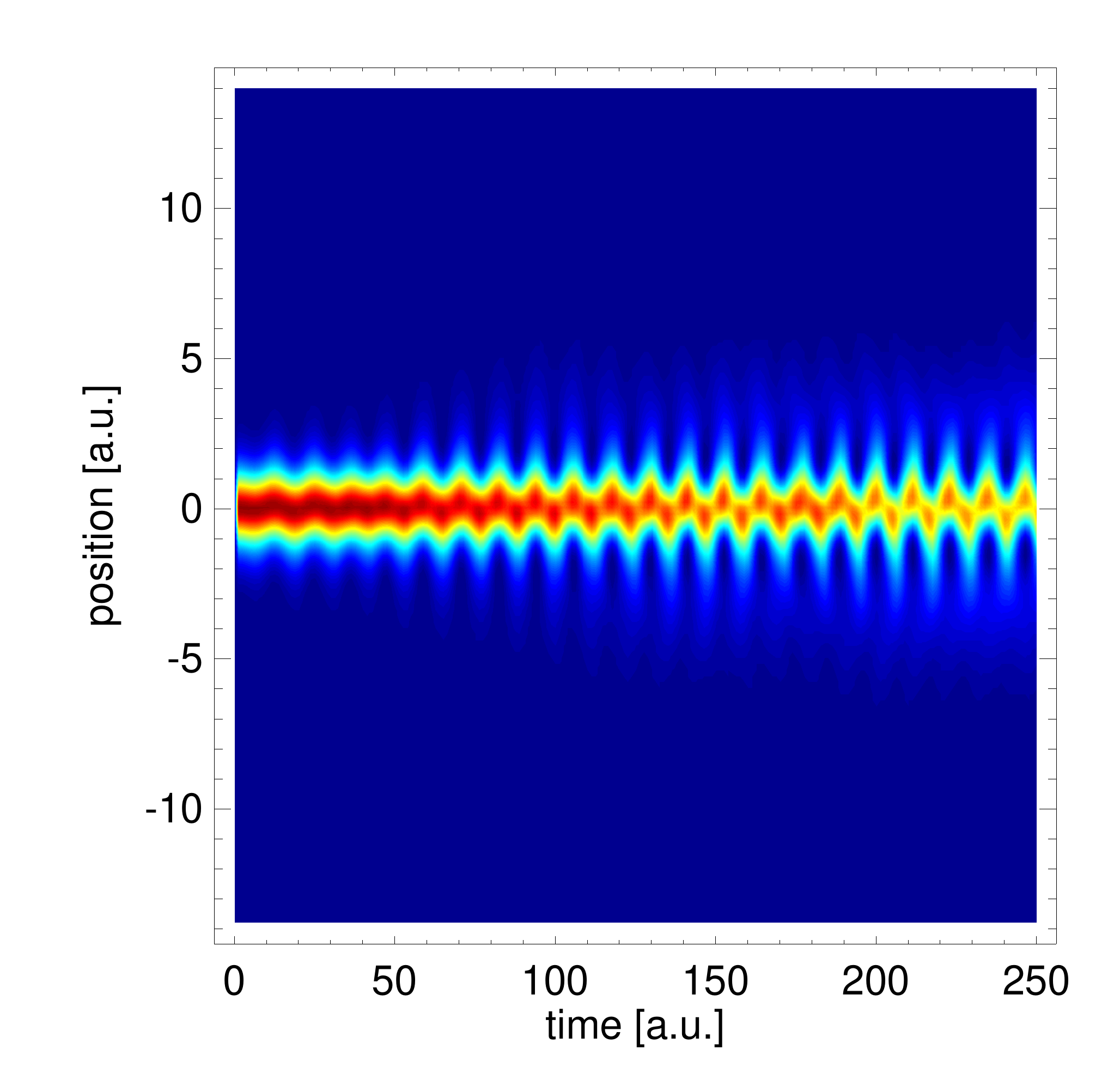}} 
  \par}
  \caption{Time-evolution of the orbital density of the first natural orbital 
   of Helium during the excitation from the ground state to the first excited state. The quiver
  motion of the electrons is nicely reflected in the density profile of the
  natural orbital.
  \label{fig:octorb}}
\end{figure}

Case(ii) - {\em optimal control}.
To study laser induced transitions in the Helium atom we add an external dipole 
laser field of the form $\hat{V}(x_1,x_2) = \left(\hat{x}_1 + \hat{x}_2\right)\,{E(t)}$ 
to the Hamiltonian $\hat{H}_{\lambda=1}$. We use standard optimal control theory (OCT) \cite{ZBR98} to find 
the optimal laser pulse with amplitude $E(t)$ which drives the atom in a finite
time-interval $[0,t_{\rm{end}}]$ from the 
ground state to the first excited state. 
The full solution of the time-dependent many-body Schr\"odinger equation shows, that the actual
transition is a mixture of both anticipated scenarios: the natural orbitals {\em as well as}
the occupation numbers change as function of time during the transition. The occupation
numbers undergo large changes which reflects the multi-reference nature of the first
excited state while the orbitals nicely reflect the quiver motion of the electrons in 
the laser field.
This is displayed in Fig.~\ref{fig:octocc} and Fig.~\ref{fig:octorb}, where
we plot optimal laser pulse, correlation entropy, occupation numbers and the orbital
density of the natural orbital with the largest occupation number.

In summary, we have presented equations of motion for the occupation numbers of the natural
spin orbitals. We have 
shown that the adiabatic extension of present ground-state 
functionals of RDMFT to the time-dependent domain yields always occupation numbers 
which stay constant in time. The exact time-evolution of the natural orbitals and
occupation numbers has been illustrated for electron-ion scattering and for the
Helium 1s-2p transition. The exact analysis shows that sizable changes in the occupation
numbers can occur during the time-evolution of the system.
Approximations beyond the ones used for static RDMFT will therefore
be neccesary to reasonably capture the time-evolution of the one-body reduced density matrix.

We would like to thank N. Helbig, I. V. Tokatly and N. N. Lathiotakis
for useful discussions.
Partial financial support by the EC Nanoquanta NoE and the
EXC!TING network is gratefully acknowledged.

\end{document}